\documentclass[aps,prb,twocolumn]{revtex4-2}
\usepackage{epsfig}
\usepackage[dvipsnames,usenames]{color}
\usepackage{color}
\usepackage{amsmath}
\usepackage{amssymb}
\usepackage{hyperref}
\usepackage{graphicx}

\usepackage{wasysym}
\usepackage{times}
\usepackage{comment}
\usepackage{array}
\usepackage{multirow}
\usepackage{tabularx}
\usepackage{float}
\usepackage[utf8]{inputenc}
\usepackage[T1]{fontenc}
\usepackage{cancel}
\usepackage{ulem}

\emergencystretch=\maxdimen
\newcommand{\bK}{{\bf K}}

\begin{document}
\title{Enhancing $d$-wave superconductivity with nearest-neighbor attraction of \\extended Hubbard model}
\author{Mi Jiang}
\affiliation{Institute of Theoretical and Applied Physics, Jiangsu Key Laboratory of Thin Films, School of Physical Science and Technology, Soochow University, Suzhou 215006, China}

\begin{abstract}
Motivated by the recent discovery of the anomalously nearest-neighbor attraction arising from the electron-phonon coupling, we quantitatively investigate the enhancing effects of this additional attractive channel on the $d$-wave SC based on dynamic cluster quantum Monte Carlo calculations of doped two-dimensional extended Hubbard model with nearest-neighbor attraction $-V$. Focusing on the range of $0<-V/t \le 2$, our simulations indicate that the dynamics of $d$-wave projected pairing interaction is attractive at all frequencies and increases with $|V|$. 
Moreover, turning on $-V$ attraction enhances the $(\pi,\pi)$ spin fluctuations but only enhances (suppresses) the charge fluctuations for small (large) momentum transfer. Thus, at $V/t=-1$ relevant to ``holon folding branch'', the charge fluctuations are insufficient to compete with $d$-wave pairing interaction strengthened by enhanced spin fluctuations. 
Our work suggest the underlying rich interplay between the spin and charge fluctuations in giving rise to the superconducting properties.
\end{abstract}

% \pacs{71.10.Fd, 71.30.+h, 02.70.Uu}
\maketitle

\section{Introduction}

The pairing mechanism mediated by virtual exchange of a bosonic mode plays the key role in overcoming the Coulomb repulsion between
electrons in order to give rise to a net attractive interaction for Cooper pairing. In conventional Bardeen-Cooper-Schrieffer (BCS) superconductors, this bosonic mechanism is realized by the retardation nature of the electron-phonon interaction~\cite{AndersonMorel1962}. Despite that there is no general consensus, there have been strong evidence that in strongly correlated superconductors such as the cuprates and heavy fermion materials, the antiferromagnetic spin fluctuations, namely the magnons, play the role of the bosonic mode. In this scenario, the minimization of the repulsive interaction due to the local Coulomb repulsion can be accomplished via the sign changing of the pairing wave function, for instance, the $d_{x^2-y^2}$-wave pair state in the cuprates~\cite{ScalapinoRMP}. 

Regarding the pairing mechanism in cuprates, there has been long debate on the role of the electron-phonon interaction and particularly its relation to superconductivity (SC). Although it is widely believed that the pure electron-electron interaction dominantly drive the Cooper pairing and the electron-phonon coupling (EPC) only plays the minor role, there has been spectroscopic evidence that the effects of strong electronic interaction and the EPC reinforce each other to drive a stronger SC in the strange-metal regime of Bi-2212~\cite{phonon1}, which indicates the possible enhancement of SC through multiple channels, for instance, the contribution from the phonon coupling, in addition to the pure electronic interaction. 
In fact, the EPC does not only manifest its importance in the enhancement of $T_c$. Most recently, comparative spectroscopic and theoretical investigation of a one-dimensional cuprate Ba$_{2-x}$Sr$_x$CuO$_{3+d}$ over a wide range of hole doping revealed the existence of an anomalously strong nearest-neighbor attraction~\cite{phonon2}, which probably originates from the EPC, in accounting for the so-called ``holon folding branch'' feature~\cite{phonon3}. 

Given the structural similarity among the cuprates, the physics with nearest-neighbor attraction of the one-dimensional material should be naturally extended to two-dimensional CuO$_2$ planes. Because how to enhance the superconducting $T_c$ is an important open question, the effects of the additional attractive channel and its interplay with the pure electron-electron interaction deserves more systematic exploration. Here we adopt an extended Hubbard model with both strong local repulsion and nearest-neighbor {\it attraction} as the minimal model. In particular, we focus on the explicit enhancement of the $d$-wave SC by the inclusion of additional strong nearest-neighbor attraction. The Hamiltonian reads as
\begin{align} \label{eq:HM}
	H = &-t\sum_{\langle ij\rangle,\sigma}
	(c^\dagger_{i\sigma}c^{\phantom\dagger}_{j\sigma}+h.c.) + U \sum_i
	n_{i\uparrow}n_{i\downarrow}\nonumber\\ &+ V \sum_{\langle
	ij\rangle,\sigma\sigma'} n_{i\sigma}n_{j\sigma'}
\end{align}
with the usual nearest-neighbor hopping $t=1$ as the unit energy scale, the on-site Coulomb repulsion $U$, and an additional nearest-neighbor Coulomb attraction $V<0$. Note that this attractive $V$ has importance difference from the conventional extended Hubbard model with repulsive $V$, which has been widely studied for the physics induced by the non-local Coulomb repulsion~\cite{Tremblay2013,TPD2015,Tremblay2016,Gull2017,Gull2018,MJ2018,Gull2019,Wehling2019,Peschke2020}. 
Regarding its superconducting properties, the consensus is that the $d$-wave pairing and the associated transition temperature are only weakly suppressed as long as the repulsive $V$ does not exceed $U/2$. This robustness is owing to the retarded nature of $d$-wave pairing to minimize the impact of non-local repulsion~\cite{Tremblay2013,Tremblay2016,MJ2018}. In the case of negative $V$, it is naively expected that attractive nearest-neighbor interactions always enhance the SC because the neighboring attraction naturally contributes the $d$-wave pairs as indicated by an early Hartree-Fock calculations~\cite{meanfield}. Conversely, the recent numerical exact diagonalization study~\cite{TPD2015} uncovered that the nearest-neighbor attractions also have thresholds above which the SC will be finally suppressed, which corrects the intuition that attractive and repulsive interactions have definitely opposite effects on SC. We emphasize that the enhanced SC explored in this work is around the moderate $0 \le |V| \le 2t$ range, which is much smaller than the threshold needed to suppress SC, to be consistent with the amplitude of the anomalously nearest-neighbor attraction $|V|\sim t$ extracted from both experimental and theoretical studies~\cite{phonon2,phonon3}.
Besides, we neglect the important but still open question of whether the pure Hubbard model at $V=0$ hosts a superconducting ground state or not~\cite{Qin,Qinreview}.

%%%%%%%%%%%%%%%%%%%%%%%%%%%%%%%%%%%%
\section{Dynamical cluster approximation}
Here we adopt the dynamical cluster approximation (DCA)~\cite{Hettler98,Maier05,code} with a continuous time auxilary field (CT-AUX) quantum Monte Carlo (QMC) cluster solver~\cite{GullCTAUX} to
numerically solve the model Eq.~\eqref{eq:HM}.
As one of various embedded-cluster methods, similar to cluster dynamical mean field theory (cDMFT), DCA maps the bulk lattice problem onto a finite cluster of size $N_c$, whose physics involving complex interactions is solved exactly by various methods e.g. QMC and exact diagonalization, while the remaining degrees of freedom are treated at the mean-field level. 
Precisely, the first Brillouin zone is divided into $N_c$ patches denoted by its center wave vector $\mathbf{K}$ surrounded by $N/N_c$ lattice wave vectors $\mathbf{k}$'s. In this way, the original lattice problem of $N$ sites is simplified to an effective $N_c$-site cluster problem by coarse graining the lattice single-particle Green's function, which is designed to converge to a cluster Green's function obtained by the cluster solver mentioned earlier~\cite{Hettler98,Maier05}. Although the inter-cluster interactions can be treated more accurately with an additional bosonic dynamic mean-field~\cite{Haule07} as adopted in the extended DMFT~\cite{Smith00}, in this work we neglect its dynamic contribution for simplicity~\cite{MJ2018}.

To achieve the goal of simulating a wide range of doping levels, most of our calculations are for smallest $N_c=2\times2$ DCA cluster to manage the sign problem of the underlying CT-AUX QMC solver~\cite{GullCTAUX,submatrix} 
down to the SC transition temperatures $T\sim T_c$. 
Despite of the small cluster size, the pairing interaction and dynamics should be fully descriptive at this level. 
In fact, the simulations with larger cluster $N_c=4\times4$ are also performed to (1) confirm the enhancing effects of the attractive $V$ while at higher temperature scale due to the QMC sign problem and (2) to investigate the competing role of spin and charge fluctuations in a finer momentum resolution. 

To investigate the superconducting, charge, and magnetic instability of a particular model Hamiltonian, one has to determine the structure of the interaction responsible for these channels. Essentially, the cluster two-particle Green's function 
\begin{align} \label{e1}
  \chi_{c\sigma\sigma'}(q,K,K') &= \int^{\beta}_0 \int^{\beta}_0 \int^{\beta}_0 \int^{\beta}_0 d\tau_1 d\tau_2 d\tau_3 d\tau_4 \nonumber \\
  & \times e^{i[(\omega_n+\nu)\tau_1 -\omega_n\tau_2 +\omega_{n'}\tau_3 -(\omega_{n'}+\nu)\tau_4]} \nonumber \\
  \times \langle \mathcal{T} & c^{\dagger}_{K+q,\sigma}(\tau_1) c^{\phantom{\dagger}}_{K\sigma}(\tau_2) c^{\dagger}_{K'\sigma'}(\tau_3) c^{\phantom{\dagger}}_{K'+q,\sigma'}(\tau_4) \rangle
\end{align}
with conventional notation $K=(\mathbf{K},i\omega_n)$, $K'=(\mathbf{K'},i\omega_{n'})$, $q=(\mathbf{q},i\nu)$ and the time-ordering operator $\mathcal{T}$ can be calculated numerically via a DCA cluster solver (CT-AUX in our case).
Then the cluster two-particle irreducible vertex $\Gamma_{c\sigma\sigma'}(q,K,K')$ can be extracted through the Bethe-Salpeter equation (BSE)
\begin{align} \label{e2}
  \chi_{c\sigma\sigma'}(q,K,K') &= \chi^0_{c\sigma\sigma'}(q,K,K') + \chi^0_{c\sigma\sigma''}(q,K,K'') \nonumber \\
  & \times \Gamma_{c\sigma''\sigma'''}(q,K'',K''') \chi_{c\sigma'''\sigma'}(q,K''',K')
\end{align}
where $\chi^0_{c\sigma\sigma'}(q,K,K')$ is the non-interacting two-particle Green's function constructed from the product of a pair of fully dressed single-particle Green's functions. The usual convention that the summation is to be made for repeated indices is adopted. 

Note that the above formalism Eqs.~(2-3) has their counterparts for the corresponding lattice quantities, whose numerical calculations are, however, impractical due to their continuous nature. Therefore, one of the key DCA assumptions is that the cluster two-particle irreducible vertex $\Gamma_c$ is used as the approximation of the desired lattice two-particle irreducible vertex $\Gamma$.

The two-particle irreducible vertex and associated BSE Eq.~\eqref{e2} can be classified according to the superconducting, charge, and magnetic channels. In this work, we are mostly interested in the particle-particle superconducting channel for the zero center-of-mass and energy.
To this aim, the superconductivity can be quantitatively displayed by the leading eigenvalues of the BSE in the particle-particle channel in the eigen-equation form~\cite{Maier06,Scalapino06}
\begin{align} \label{eq:BSE}
    -\frac{T}{N_c}\sum_{K'}
	\Gamma^{pp}(K,K')
	\bar{\chi}_0^{pp}(K')\phi_\alpha(K') =\lambda_\alpha(T) \phi_\alpha(K)
\end{align}
where $\Gamma^{pp}(K,K')$ denotes the irreducible particle-particle vertex of the effective cluster problem with the cluster momenta $\bK$ and Matsubara frequencies $\omega_n=(2n+1)\pi T$. Note that the spin indices are neglected for simplicity. Besides, for the superconducting channel, $q=(\mathbf{q},i\nu)=0$ is assumed since our focus in this work is the even-frequency even-parity (spin singlet) $d$-wave pairing tendency~\cite{Maier06,Scalapino06}.
The coarse-grained bare particle-particle susceptibility
\begin{align}\label{eq:chipp}
	\bar{\chi}^{pp}_0(K) = \frac{N_c}{N}\sum_{k'}G(K+k')G(-K-k')
\end{align}
is obtained via the dressed single-particle Green's function $G(k)\equiv G({\bf k},i\omega_n) =
[i\omega_n+\mu-\varepsilon_{\bf k}-\Sigma({\bf K},i\omega_n)]^{-1}$, where $\mathbf{k}$ belongs to the DCA patch surrounding the cluster momentum $\mathbf{K}$, $\mu$ the chemical potential, $\varepsilon_{\bf k}=-2t(\cos k_x+\cos k_y)$ the
dispersion relation, and $\Sigma({\bf K},i\omega_n)$ the cluster self-energy. In practice, we usually choose 16 discrete points for both the positive and negative fermionic Matsubara frequency $\omega_n=(2n+1)\pi T$ mesh for measuring the four-point quantities like two-particle Green's functions and irreducible vertices. Therefore, the BSE Eq.~\eqref{eq:BSE} reduces to an eigenvalue problem of a matrix of size $(32N_c)\times (32N_c)$.

The eigenvalue $\lambda_\alpha(T)$ gives the pairing tendency of the superconducting channel; while the symmetry of the corresponding superconducting state is manifested by the momentum and frequency dependence of the eigenvector $\phi_\alpha({\bf K},i\omega_n)$. 
Note that the magnitude of $\lambda_\alpha(T)$ denotes the strength of the normal state pairing correlations. Accordingly, the spatial, frequency, and more generally orbital dependence of the eigenvector $\phi_\alpha({\bf K},i\omega_n)$ can be viewed as the normal state analog of the superconducting gap to reflect the structure of the pairing interaction~\cite{Maier06,Scalapino06}.
The superconducting $T_c$ is extracted via the temperature where the leading eigenvalue of Eq.~\eqref{eq:BSE} $\lambda(T_c)=1$.
As expected for the extended Hubbard model, the leading pairing symmetry occurs for the $d$-wave channel with momentum structure $\cos K_x - \cos K_y$ so that we are only concerned in the leading eigenvalues $\lambda_d$ and associated $\phi_d({\bf K},i\omega_n)$. 

As discussed by Scalapino~\cite{Scalapino06}, the two-particle irreducible particle-particle vertex $\Gamma^{pp}$ as the pairing interaction is connected to the four-point vertex function, which also contains information of the irreducible particle-hole vertex $\Gamma^{ph}$ in magnetic and charge channels. Thus, the pairing interaction $\Gamma^{pp}$ has intrinsic relation to these particle-hole channels $\Gamma^{ph}$. 
In fact, the dominant contribution on the $d$-wave pairing interaction has been shown to arise from the spin-one ($S=1$) particle-hole exchange~\cite{Maier06,Scalapino06}.
Therefore, in this work we also extract the irreducible particle-hole vertex $\Gamma^{ph}$ in the magnetic and charge channels of the effective cluster problem respectively from Eq.~\eqref{e2} and thereby we have the BSE in the eigen-equation form similar to Eq.~\eqref{eq:BSE} but with coarse-grained bare particle-hole susceptibility
\begin{align}\label{eq:chiph}
	\bar{\chi}^{ph}_{0}(q,K,K') = \delta_{KK'} \frac{N_c}{N} \sum_{k'}  G(K+k') G(K+k'+q)
\end{align}
The corresponding eigenvalues for the particle-hole channels reflect the magnetic and charge instabilities, e.g. spin and charge density waves.

In this work we are only interested in the case of zero frequency transfer ($i\nu=0$) similar to the particle-particle superconducting channel Eq.~\eqref{eq:BSE}. Note, however, that we keep the momentum transfer to calculate $\mathbf{q}$-dependent lattice susceptibilities, which can be obtained by the coarse-grained two-particle Green's function $\bar{\chi}^{ph}(q,K,K')$ (instead of cluster quantities that result in cluster susceptibilities), which is in turn calculated via the coarse-grained BSE transformed from Eq.~\eqref{e2} as
\begin{align}\label{}
	[\bar{\chi}^{ph}(q,K,K')]^{-1} = [\bar{\chi}^{ph}_{0}(q,K,K')]^{-1} - \Gamma^{ph}(q,K,K')
\end{align}
Then our interested magnetic (s) and charge (c) lattice susceptibilities $\chi_{s,c}(q,T)$ can be deduced as
\begin{align}\label{chics}
	\chi_{s,c}(q,T) = \frac{T^2}{N_c^2} \sum_{K,K'} \bar{\chi}^{ph}(q,K,K')
\end{align}
We refer the readers to Ref.~\cite{dca2001} for more details of the DCA formalism of the calculations of two-particle quantities.

%%%%%%%%%%%%%%%%%%%%%%%%%%%%%%%%%%%%%%%%%%%%%%%%%%%%%%%%%%%%%
\section{Results}
\begin{figure}
\psfig{figure=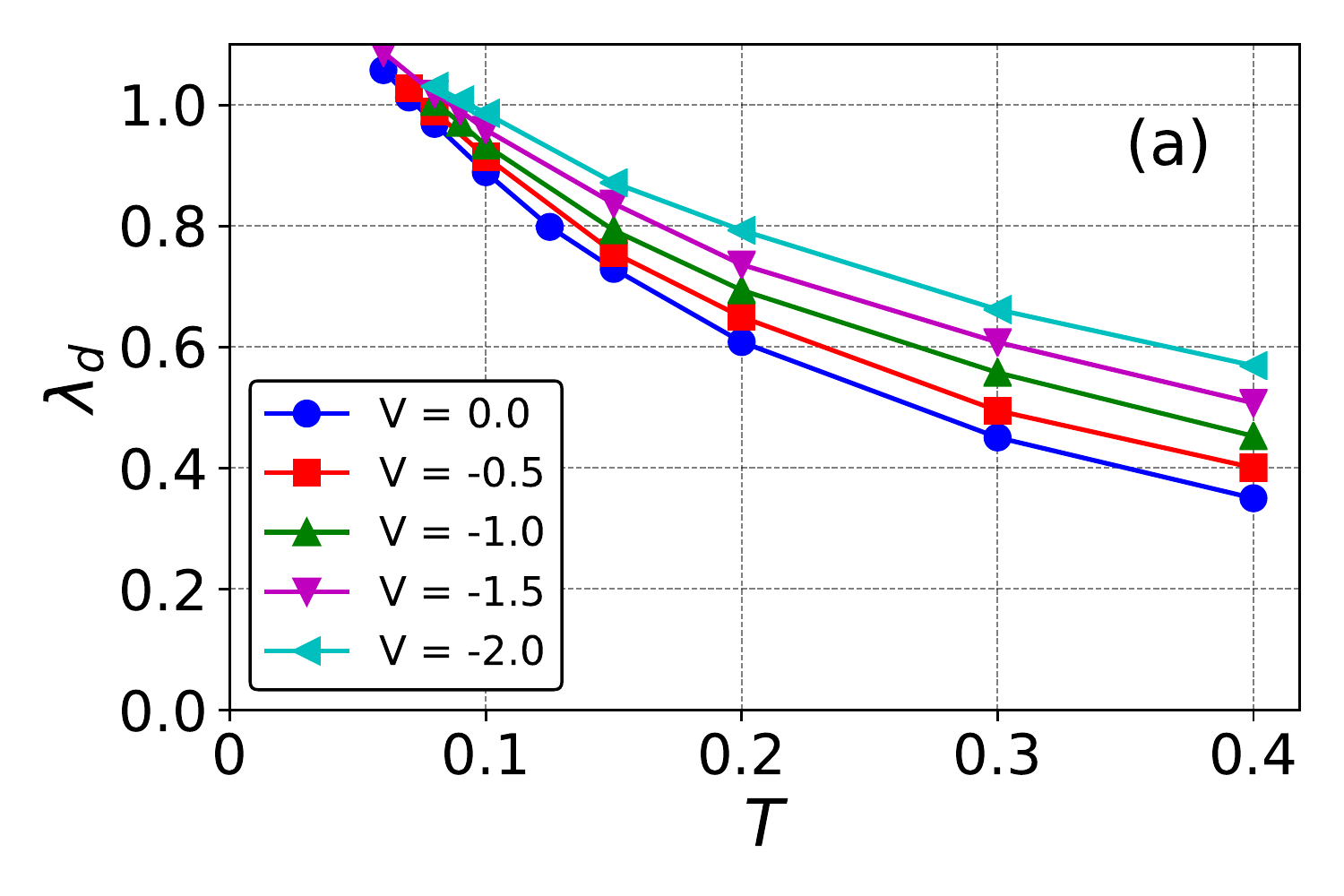,height=5.5cm,width=8.5cm,angle=0,clip} \\
\psfig{figure=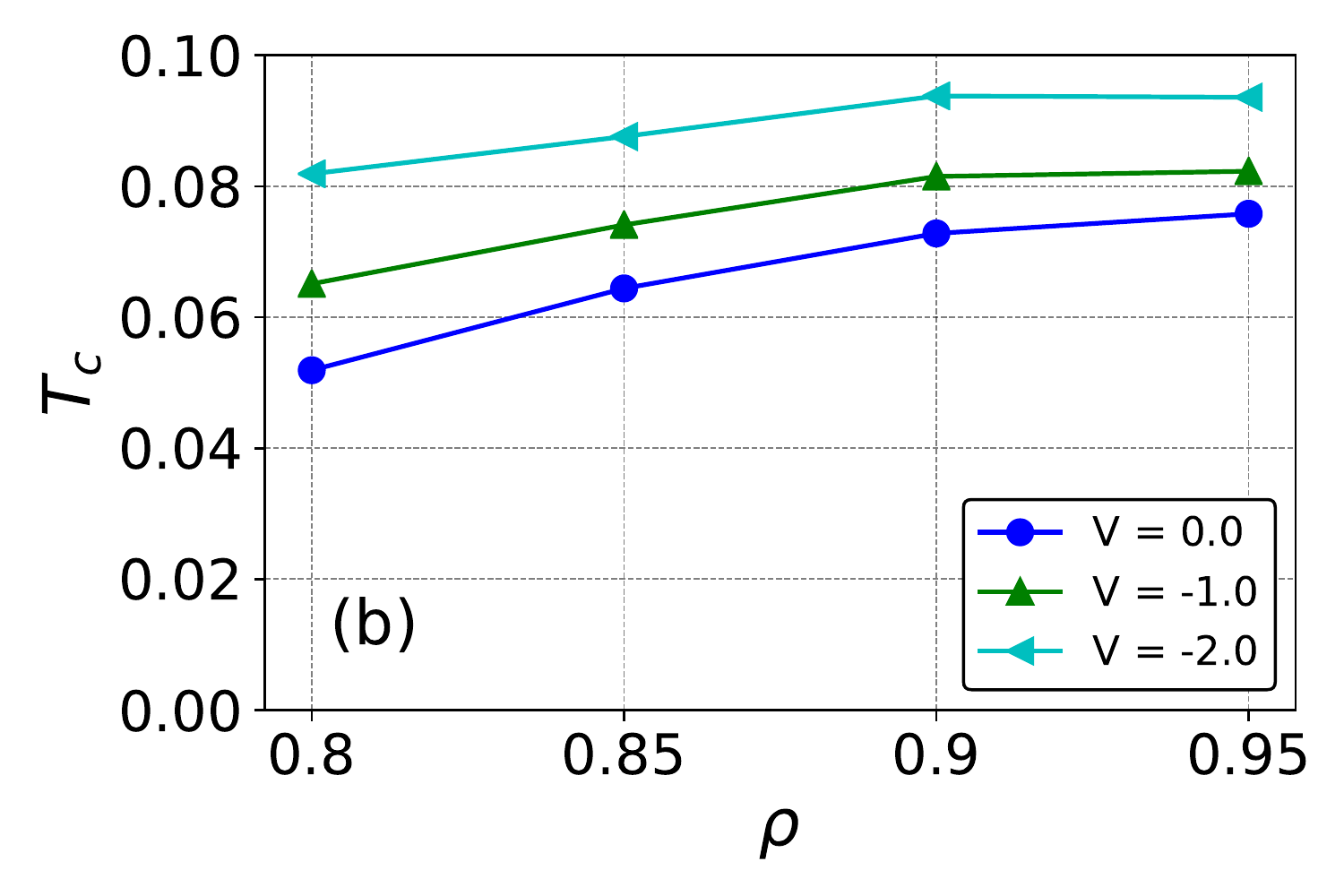,height=5.5cm,width=8.5cm,angle=0,clip} \\
\psfig{figure=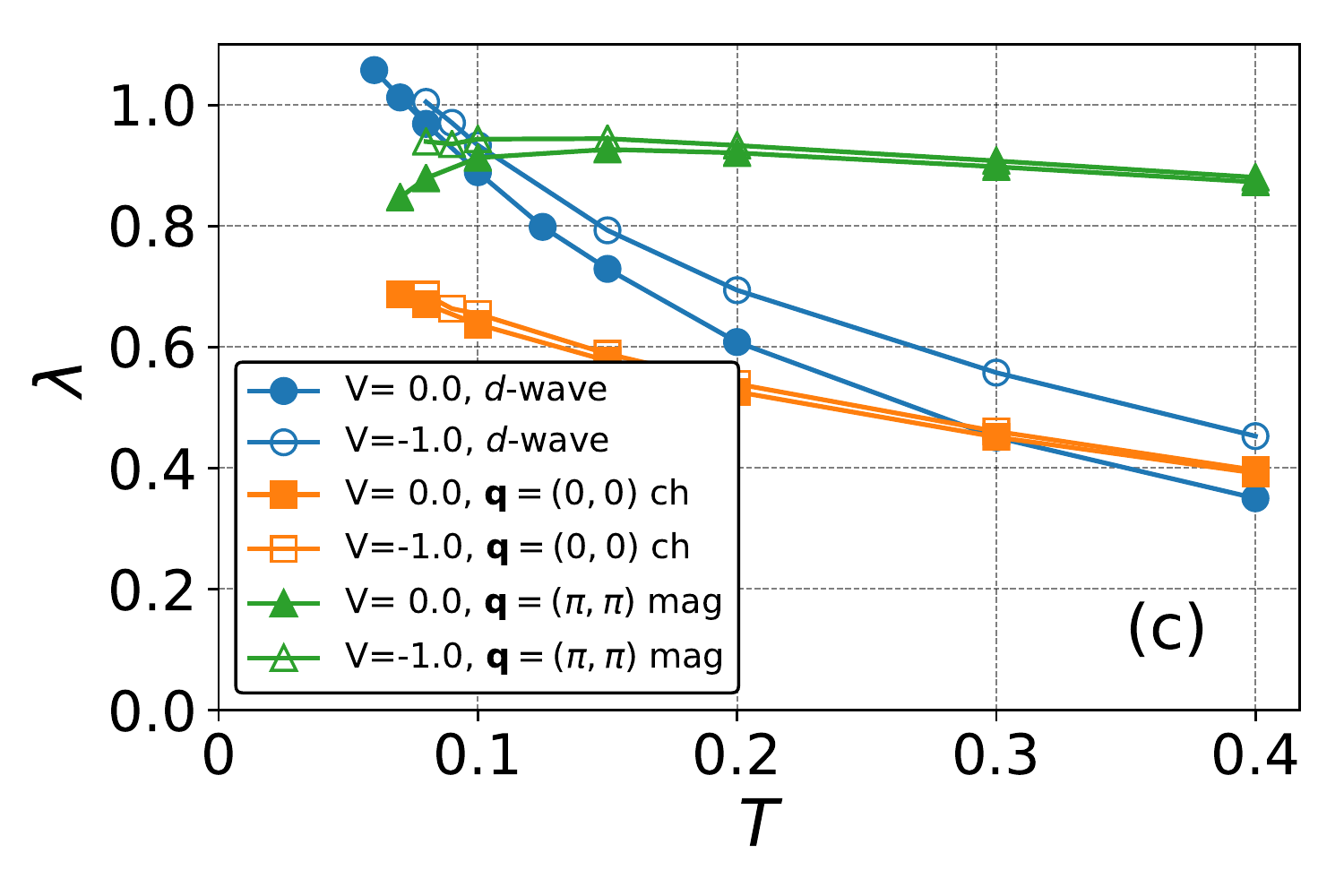,height=5.5cm,width=8.5cm,angle=0,clip} 
\caption{(a) Temperature dependence of the leading ($d_{x^2-y^2}$-wave) eigenvalue $\lambda_ {d}(T)$ of BSE Eq.~\eqref{eq:BSE} in the particle-particle channel at $U/t=7$ and $\langle n\rangle=0.9$; (b) The filling $\rho$ dependence of the $d$-wave superconducting $T_{c}$ extracted from
$\lambda_d(T_c)=1$ reveals the enhancing effect of nearest-neighbor attraction; (c) Comparison of the leading eigenvalues for $d$-wave superconducting, $\mathbf{q}=(\pi,\pi)$ antiferromagnetic, and $\mathbf{q}=(0,0)$ charge channels.}
\label{lambda}
\end{figure}

We first illustrate the temperature dependence of the leading $d$-wave eigenvalue $\lambda_d(T)$ for different $V$ at fixed filling $\rho=0.9$ in Fig.~\ref{lambda}(a).
Apparently, the nearest-neighbor attractive $V$ leads to the increase of $\lambda_{d}(T)$ as the evidence that the $d$-wave pairing tendency can be enhanced. 
To clearly show the enhancing effects of $V$, Fig.~\ref{lambda}(b) displays the dependence of $T_c$ extracted via $\lambda_{d}(T_c)=1$ on the filling. One can see that approximately 10-15\% enhancement of $T_c$ with finite attraction is a general feature for all fillings considered here. Compared with the impact of repulsive $V$ on the $d$-wave pairing~\cite{MJ2018}, the variation of $T_c$ with $\pm V$ is not exactly symmetric over $V=0$ but has roughly the same scale. Therefore, in this sense, the $d$-wave pairing has similar robustness against the additional channel of attractive interaction from nearest-neighbor $-V$. In other words, the anomalously nearest-neighbor attraction $|V|\sim t$ extracted from both experimental and theoretical studies to account for the ``holon folding branch''~\cite{phonon2,phonon3} does not have significant effects on the $d$-wave superconducting $T_c$.

\begin{figure}
\psfig{figure=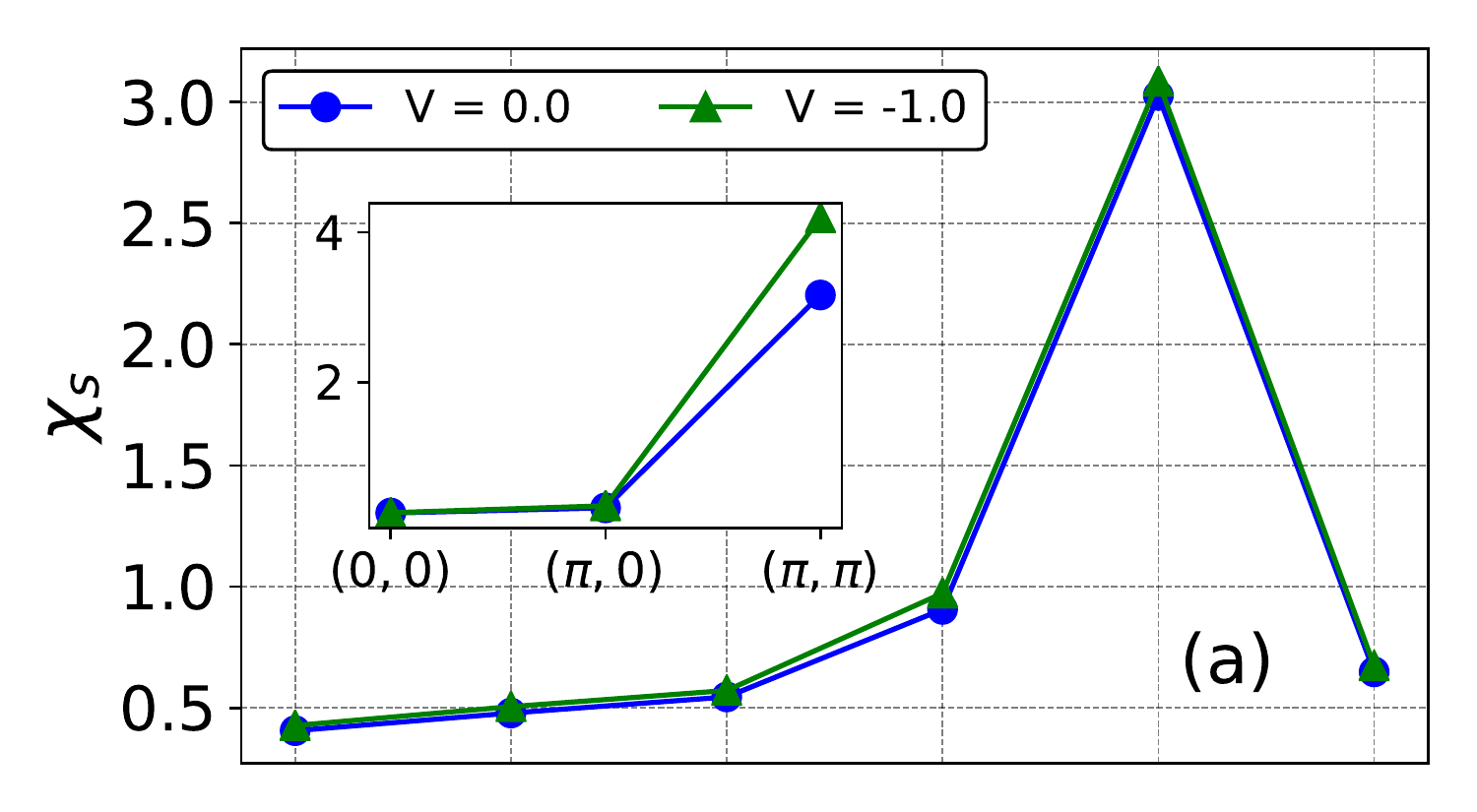,height=4.2cm,width=8.0cm,angle=0,clip} \\
\psfig{figure=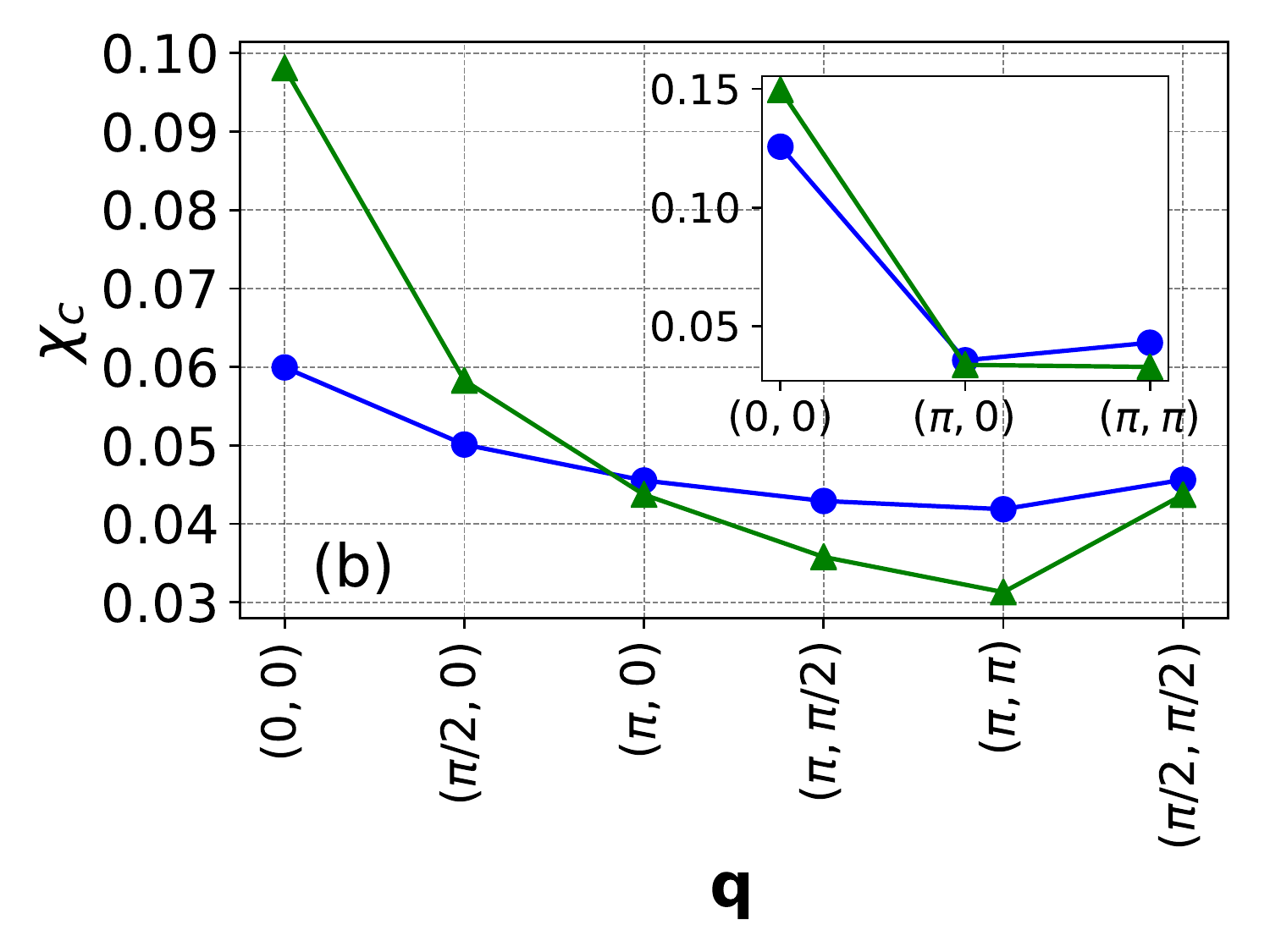,height=6.0cm,width=8.0cm,angle=0,clip} 
\caption{The momentum transfer $\mathbf{q}$ dependence of the DCA lattice (a) spin and (b) charge susceptibilities, $\chi_s$ and $\chi_c$ respectively, for DCA cluster $N_c=4\times 4, T/t=0.3, U/t=7, \langle n\rangle=0.9$. Turning on $-V$ attraction slightly (because of relatively high temperature) enhances the spin fluctuations at all $\mathbf{q}$'s but only enhances (suppresses) the charge fluctuations for small (large) $\mathbf{q}$. The inset shows the results for smaller cluster $N_c=4$ but lower $T/t=0.1$ to illustrate the enhancement of $\chi_s(\pi,\pi)$.}
\label{chi}
\end{figure}

Besides the $d$-wave superconductivity, the (extended) Hubbard model can support other instabilities like spin and charge density waves (SDW/CDW)~\cite{Tremblay2013,TPD2015,Tremblay2016,Gull2017,Gull2018,MJ2018,Gull2019,Wehling2019,Peschke2020}. In particular, the additional nearest-neighbor attraction is natural to host the charge ordering instability. Therefore, to explore these instability apart from superconductivity, Fig.~\ref{lambda}(c) compares the temperature evolution of the leading eigenvalues for $d$-wave superconducting, $\mathbf{q}=(\pi,\pi)$ antiferromagnetic, and $\mathbf{q}=(0,0)$ charge channels. The choice of these two particular $\mathbf{q}$'s is motivated by Fig.~\ref{chi}, which indicates that the leading lattice magnetic and charge susceptibilities locate at these two specific momentum transfers.

On the one hand, apparently, the dominant instability at low enough temperature is $d$-wave superconducting while at higher temperatures the antiferromagnetic ordering instability exceeds the pairing one~\cite{Maier06}. On the other hand, the expected charge ordering instability is always the subleading one since our interested nearest-neighbor attraction $V/t=-1$ is still weak to suppress the leading $d$-wave pairing instability to induce the desired charge ordering.

\begin{figure}
\psfig{figure=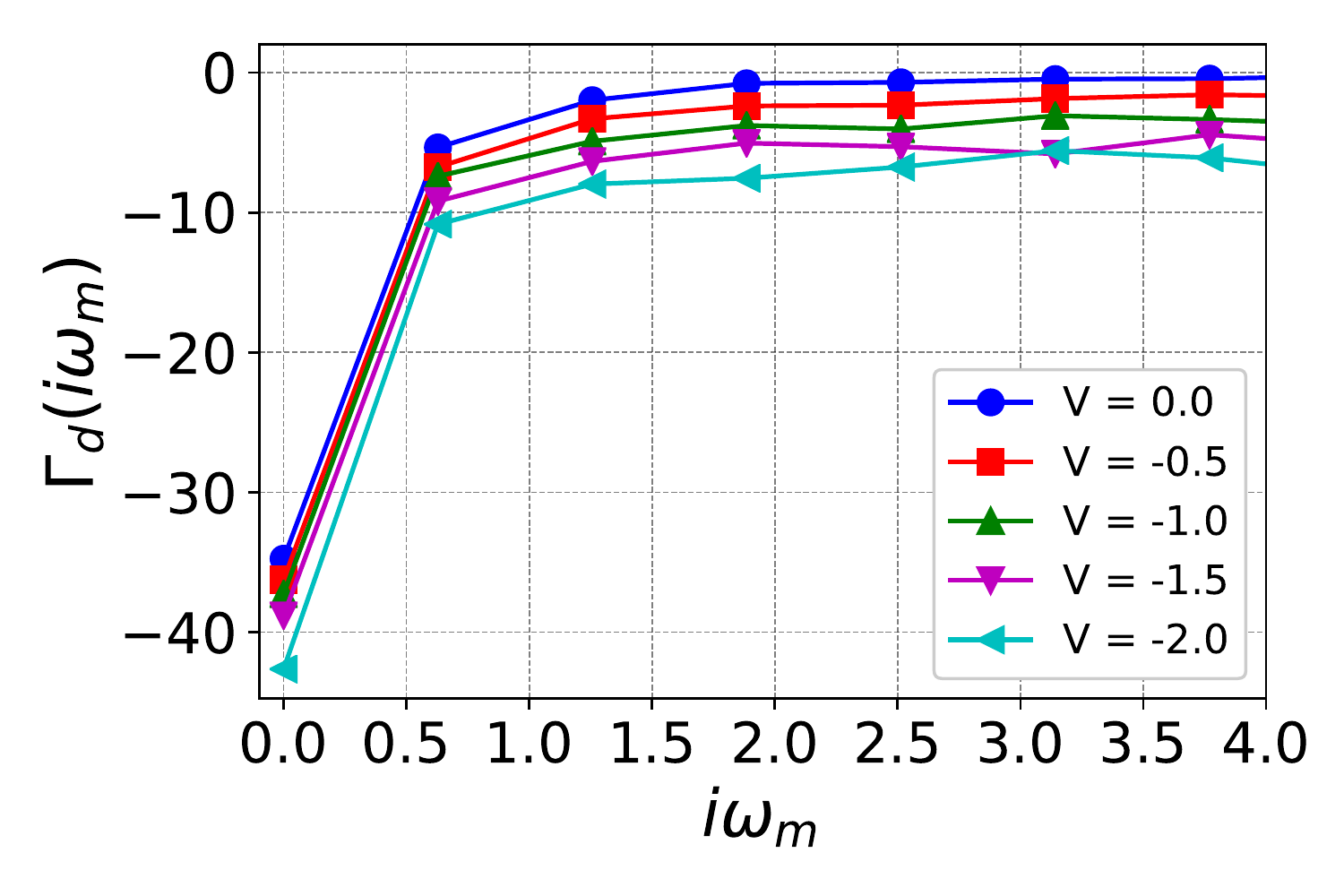,height=5.5cm,width=8.5cm,angle=0,clip}
\caption{The $d$-wave projected irreducible particle-particle
vertex $\Gamma_{d}(i\omega_{m})$ for different attraction $V$ at $U/t=7$ and $\langle
    n\rangle=0.9$ at $T/t=0.1$. $\Gamma_d$ is attractive at all frequencies.}
\label{Vd}
\end{figure}

To have a better understanding of the pairing interaction and its variation with the additional
nearest-neighbor attraction, we resort to the $d$-wave projected dynamical pairing interaction 
\begin{equation}\label{}
  \Gamma_{d}(i\omega_m) = \frac{\sum\limits_{
  \mathbf{K,K'}} g_d(
  \mathbf{K})
  \Gamma^{pp}(\mathbf{K},i\omega_{n} , \mathbf{K'},i\omega_{n'})
  g_d(\mathbf{K'})}{\sum\limits_{\mathbf{K}} g_d^{2}(\mathbf{K})}
\end{equation}
and its dependence on the bosonic Matsubara frequency $\omega_m = \omega_n-\omega_{n'}$, where $g_d( \mathbf{K})=\cos K_{x} - \cos K_{y}$ gives the $d$-wave projection factor, and the fixed $\omega_{n'}=\pi T$ is adopted. 
As shown in Fig.~\ref{Vd}, the pairing interaction $\Gamma_d(i\omega_m)$ is attractive (negative) in all cases and the finite additional $-V$ naturally strengthens the attractive interaction favoring neighboring spin configuration. At high frequencies, $\Gamma^{pp}$ approaches to the bare interaction $V({\bf q}=\bK- \bK')$, which is the Fourier transform of the nearest-neighbor interaction $V$. As confirmed in Fig.~\ref{Vd}, for our 2$\times$2 cluster, we have 
$\Gamma_d(i\omega_m) \sim 4V$ at large $i\omega_m$. Different from the repulsive $V$ cases, where $\Gamma_d(i\omega_m)$ is only attractive at low frequencies but repulsive at high enough frequencies so that the effective attraction arises from the low frequency components of $\Gamma_d(i\omega_m)$~\cite{MJ2018}, the persistent attractive nature of $\Gamma_d(i\omega_m)$ reflects the cooperation between the original effective attraction from pure local Coulomb repulsion and the additional nearest-neighbor attraction.

\begin{figure}
\psfig{figure=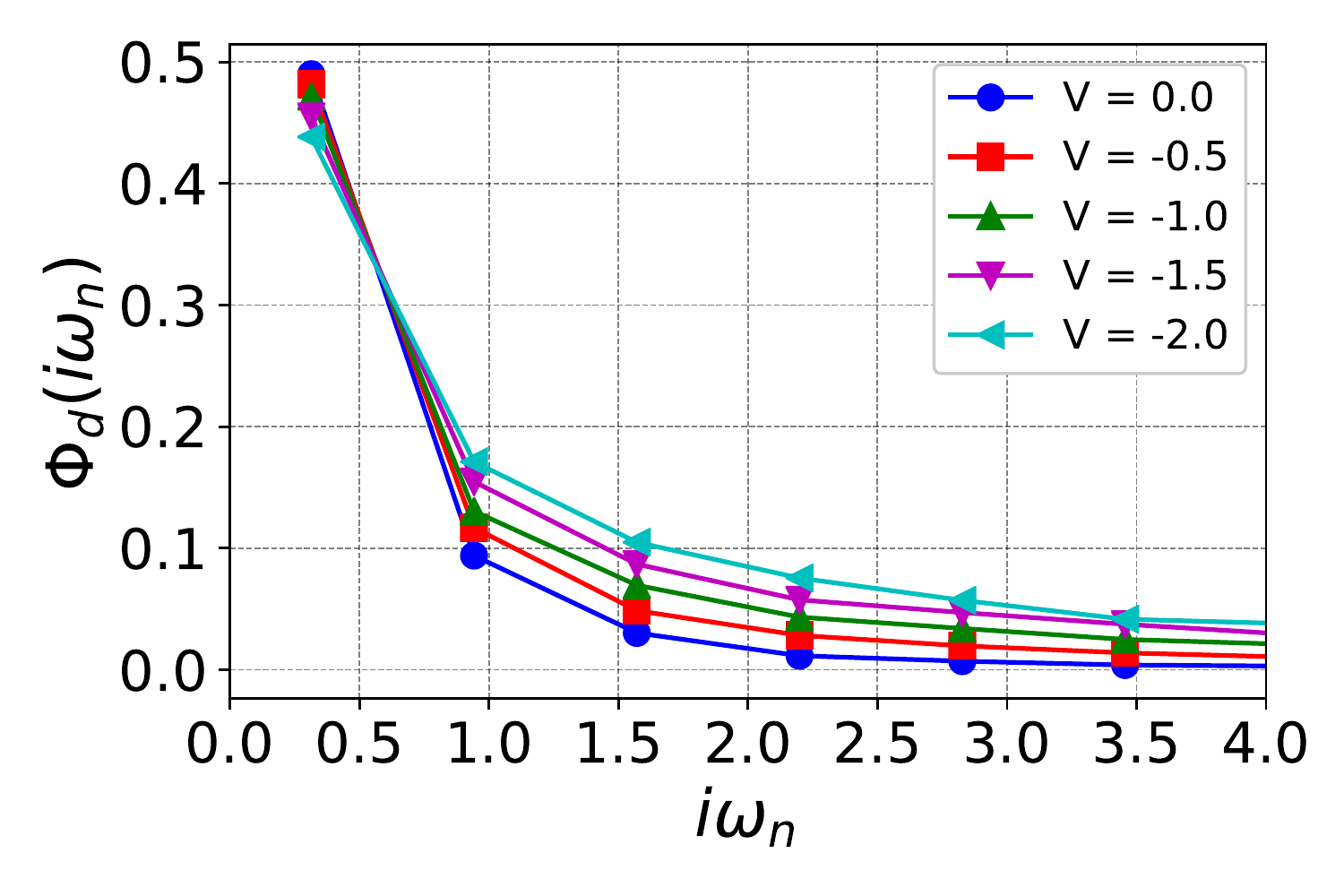,height=5.5cm,width=8.5cm,angle=0,clip}
\\ \caption{The leading $d$-wave eigenfunction $\phi_d({\bf K}=(\pi,0),i\omega_n)$ of BSE Eq.~\eqref{eq:BSE} for different $V$ at $U/t=7$, $T/t=0.1$ and $\langle n\rangle=0.9$. The retardation of $\phi_d$ becomes stronger with increasing $|V|$.} 
\label{eigenvector}
\end{figure}

The attractive feature and retardation nature of the $d$-wave pairing interaction can be reflected via the leading $d$-wave eigenvector $\phi_d({\bf K},i\omega_n)$ of Eq.(2), whose frequency dependence is shown in Fig.~\ref{eigenvector} at ${\bf K}=(\pi,0)$ and $T=0.1$ for varying $V$. For all cases, $\phi_d$ falls to zero with a retardation characteristic frequency scale, which mirrors the pairing interaction in Fig.~\ref{Vd}. 

Although the linear change of $\Gamma_d(i\omega_m)$ in Fig.~\ref{Vd} looks $i\omega_m$ independent, the decisive factor in the BSE Eq.~\eqref{eq:BSE} is the ratio between $\Gamma_d$ at different $V$, which indeed strongly depends on the bosonic Matsubara frequency $i\omega_m$.
Besides, the BSE also involves the coarse-grained bare two-particle susceptibility $\bar{\chi}^{pp}_0$, whose ratio between the values at different $V$ is $i\omega_n$-dependent as well. Therefore, it is not surprising that the variation of eigenvectors $\phi_d$ with $V$ strongly depend on $i\omega_n$ instead of simple linear change.
In this mathematical sense, the relation between $\Gamma_d$ and $\phi_d$ can be complex. However, it is physically plausible that $\phi_d$ becomes more retarded because of the additional nearest-neighbor attraction, which is similar to the phonon mediation induced retardation in conventional superconductors.
This is confirmed by the gradually increasing  frequency scale of $\phi_d$'s decaying. This might also hint that at sufficiently strong attractive $V$, the leading eigenvector may lose the $d$-wave character, namely that the $d$-wave SC would be finally destroyed and replaced by the competing charge orders~\cite{TPD2015}.

%%%%%%%%%%%%%%%%%%%%%%%%%%%%%%%%%%%%%%%%%%%%%%%%%

As mentioned earlier, the pairing interaction $\Gamma^{pp}$ has intrinsic relation to the irreducible particle-hole vertex $\Gamma^{ph}$. Given that the magnetic channel plays the central role in mediating the $d$-wave pairing and also the additional nearest-neighbor interactions, either repulsive~\cite{Tremblay2013,Tremblay2016,MJ2018} or attractive~\cite{TPD2015}, favor charge ordering, we calculate the zero frequency DCA lattice spin (s) and charge (c) susceptibilities via Eq.~\ref{chics}, whose dependence on the momentum transfer $\mathbf{q}$ are shown in Fig.~\ref{chi}'s main parts for DCA cluster $N_c=4\times4$ and $T/t=0.3, \langle n\rangle=0.9$. 
At $V=0$, the magnetic susceptibility $\chi_s$ peaks at $\mathbf{q}=(\pi,\pi)$ as expected for the repulsive Hubbard model on square lattice, which is consistent with the scenario that the antiferromagnetic fluctuations mediate the $d$-wave pairing~\cite{Scalapino06,Maier06}. 
As turning on $V/t=-1$, $\chi_s$ exhibits tiny increase at all $\mathbf{q}$'s, which is due to the relatively high temperature $T/t=0.3$ to compromise with the severe sign problem at lower temperature for large DCA cluster $N_c=4\times4$. The complementary inset of Fig.~\ref{chi}(a) explicitly shows the increase of $\chi_s$ at $\mathbf{q}=(\pi,\pi)$, which is consistent with the increase of the eigenvalues in the magnetic channel (green line in Fig.~\ref{lambda}(c)). 
Apparently, the common peak structure of $\chi_s$ at finite $V$, namely the spin fluctuation is strongest at large momentum transfer, can enhance the $d$-wave pairing interaction and in turn push up $T_c$.
Hence, the physical picture in terms of the mediating role of spin fluctuations in SC is the same as the system without $V$. In addition, owing to the additional attraction, the nearest-neighbor spin configurations are favored to be compatible with the $d$-wave SC.

Compared with the behavior of $\chi_s$, the charge susceptibility $\chi_c$ shows more nontrivial features. In particular, turning on $-V$ attraction enhances the charge fluctuation at small momentum transfer e.g. $\mathbf{q}=(0,0), (\pi/2,0)$ instead of $\mathbf{q}=(\pi,\pi)$ expected for repulsive $V$ interaction~\cite{MJ2018}. The inset of Fig.~\ref{chi}(b) at $N_c=4, T/t=0.1$ has the similar variation with $\mathbf{q}$. The favored nearest-neighboring charges are prone to enhance the small $\mathbf{q}$ charge fluctuations, which coexists with the $d$-wave pairing favored by $\mathbf{q}=(\pi,\pi)$ spin fluctuations. Therefore, both attractive and repulsive interactions favor charge fluctuations but at different wave vectors, which is
reminiscent of the previous exact diagonalization investigation based on the spin and charge structure factors of extended Hubbard model~\cite{TPD2015}. Apparently, as $|V|$ exceeds some threshold, the charge fluctuations would exceed the magnetic fluctuations and finally destroy the $d$-wave SC. Summarizing Fig.~\ref{chi}, at our interested moderate $V/t=-1$, the enhanced SC originates from the enhanced spin fluctuation at large momentum transfer $\mathbf{q}=(\pi,\pi)$, where the associated charge fluctuations are suppressed, namely the charge ordering tendency at $V/t=-1$ is insufficient to suppress SC, whose impact can only manifest itself at much large $V$ attraction to host the charge order at small $\mathbf{q}$.

\section{Summary}
In conclusion, we adopted dynamic cluster quantum Monte Carlo calculations of the extended Hubbard model with nearest-neighbor attraction to study the impact of the additional attractive channel on the $d$-wave SC. In particular, we focus on the attractive interaction with amplitude $|V|\sim t$, which is motivated by the recent discovery of an anomalously strong nearest-neighbor attraction probably arising from the electron-phonon couplings~\cite{phonon2,phonon3}.

It is found that the additional $-V$ enhances the $d$-wave SC and the variation of $T_c$ with $\pm V$ has roughly the same scale, which confirms the expectation that the repulsive (attractive) $V$ suppresses (enhances) the SC before it is ultimately destroyed by sufficiently large $|V|$~\cite{TPD2015}.
Distinct from the case of repulsive $V$, the $d$-wave projected pairing interaction $\Gamma_d$ is attractive at all frequencies and its amplitude increases with $|V|$ and thereby favors the $d$-wave SC. Reflecting the behavior of $\Gamma_d$, the $d$-wave eigenfunction $\phi_d$ of the BSE in the particle-particle channel falls to zero with a characteristic frequency scale, which increases with $|V|$ indicating the stronger retardation induced by the additional $-V$. 
Furthermore, the examination of the momentum transfer $\mathbf{q}$-resolved spin and charge susceptibilities indicates that, on the one hand,
$(\pi,\pi)$ spin fluctuations become stronger at $V/t=-1$ compared with the case at $V=0$; on the other hand, $-V$ attraction only enhances the charge fluctuations for small momentum transfer instead of large wave vector e.g. $(\pi,\pi)$. Thus, at the stage of our interested $V/t=-1$ relevant to ``holon folding branch''~\cite{phonon2,phonon3}, the charge fluctuations of much smaller amplitude than its magnetic counterpart are insufficient to compete with the $d$-wave pairing interaction. 

Our presented work provides complemental knowledge on the extensively studied extended Hubbard model, especially on quantitative examination of the role of the additional nearest-neighbor attraction uncovered recently. These results suggest the underlying rich interplay between the spin and charge fluctuations in giving rise to the superconducting properties.

\section*{Acknowledgments}
We acknowledge useful discussion with Zhenzhong Shi on the cuprates and the assistance of Peizhi Mai for the code testing of related projects.
This work was supported by the startup fund from Soochow University and Priority Academic Program Development (PAPD) of Jiangsu Higher Education Institutions.

%%%%%%%%%%%%%%%%%%%%%%%%%%%%%%%%%%%%%%%%%%%%%%%%%%%%%%%%%%%%%%%%%%

\end{document}